\documentclass[twocolumn,showpacs,preprintnumbers,amsmath,amssymb,prl,superscriptaddress]{revtex4-1}
\usepackage{bbm}
\usepackage{mathrsfs}
\usepackage{graphicx}
\usepackage{dcolumn}
\usepackage{bm}
\usepackage{amsmath}
\usepackage{amsfonts}
\usepackage{color}
\usepackage[colorlinks=true,citecolor=blue,anchorcolor=cyan]{hyperref}
\usepackage{floatrow}

\begin{document}

\title{Dynamical magnetoelectric coupling in axion insulator thin films}
\author{Zhaochen Liu}
\affiliation{State Key Laboratory of Surface Physics and Department of Physics, Fudan University, Shanghai 200433, China}
\author{Jiang Xiao}
\affiliation{State Key Laboratory of Surface Physics and Department of Physics, Fudan University, Shanghai 200433, China}
\affiliation{Institute for Nanoelectronic Devices and Quantum Computing, Fudan University, Shanghai 200433, China}
\author{Jing Wang}
\thanks{wjingphys@fudan.edu.cn}
\affiliation{State Key Laboratory of Surface Physics and Department of Physics, Fudan University, Shanghai 200433, China}
\affiliation{Institute for Nanoelectronic Devices and Quantum Computing, Fudan University, Shanghai 200433, China}

\begin{abstract}
Axion insulator is an exotic magnetic topological insulator with zero Chern number but a nonzero quantized Chern-Simons magnetoelectric coupling. A conclusive experimental evidence for axion insulators is still lacking due to the small signal of topological magnetoelectric effect (TME). Here we show that the dynamical magnetoelectric coupling can be induced by the \emph{out-of-plane} surface magnetization dynamics in axion insulator thin films, which further generates a polarization current in the presence of an external magnetic field in the same direction. Such a current is finite in the bulk and increases as the film thickness $d$ decreases, in opposite to TME current which decreases as $d$ decreases. Remarkably, the current in thin films at magnetic resonance is at least ten times larger than that of TME, and thus may serve as a smoking gun signature for axion insulators.
\end{abstract}

\date{\today}


\maketitle

The search for new topological phenomena has become an important goal in condensed matter physics~\cite{thouless1998,hasan2010,qi2011}. The intricate interplay between topology and magnetism could generate a variety of exotic quantum states~\cite{tokura2019,wang2017c}. One interesting example is axion insulators (AI), which are magnetic topological insulators (TI) with zero Chern number but a nonzero \emph{quantized} Chern-Simons magnetoelectric coupling~\cite{qi2008,essin2009,coh2011,nomura2011,turner2012,wan2012,morimoto2015,wang2015b,li2010,mogi2017,mogi2017a,xiao2018,grauer2017,varnava2018,allen2019,zhang2019,gong2019,gui2019,xu2019,chowdhury2019,wieder2018,liu2020,deng2020,varnava2020}. The simplest AI is obtained in three-dimensional (3D) TIs with a surface gap induced by a hedgehog magnetization while preserving the bulk gap~\cite{qi2008,wang2015b}. The unique signature of AI is the topological magnetoelectric effect (TME)~\cite{qi2008,karch2009,mulligan2013,rosenow2017}, where a quantized polarization is induced by a parallel magnetic field. Such an electromagnetic response is described by the topological $\theta$ term $\mathcal{L}_{\theta}=(\theta/2\pi)(e^2/h)\mathbf{E}\cdot\mathbf{B}$~\cite{qi2008,wilczek1987}, together with the ordinary Maxwell Lagrangian. Here $\mathbf{E}$ and $\mathbf{B}$ are the conventional electromagnetic fields inside the insulator, $e$ is the elementary charge, $h$ is Plank's constant, $\theta$ is the dimensionless pseudoscalar axion field~\cite{peccei1977}. From the effective action with an open boundary condition, $\theta=\pi$ in AI describes a half-quantized surface anomalous Hall conductance, which is the physical origin of TME and leads to the image magnetic monopole~\cite{qi2009b} and topological magneto-optical effect~\cite{okada2016,wul2016,dziom2017}. However, no experimental confirmation of TME has been achieved due to the small TME current. AI also exhibits the zero Hall resistance with a large longitudinal resistance~\cite{wang2015b}, which has been experimentally observed in AI candidates such as ferromagnet-TI-ferromagnet (FM-TI-FM) heterostructure~\cite{mogi2017,mogi2017a,xiao2018} and MnBi$_2$Te$_4$~\cite{liu2020,deng2020}.  However, the zero Hall resistance can also exists in trivial insulators and is not conclusive. Therefore, seeking a testable transport signature for AI is still an open question.

One of the most intriguing physical phenomena driven by the topological term is the electromagnetic effect via the dynamics of $\theta$ (i.e., $\partial_t\theta$). So far, axion polariton~\cite{li2010} and axion instability~\cite{ooguri2012} have been proposed under a nonzero $\partial_t\theta$, whose dynamics is caused by magnetic fluctuations in bulk materials with breaking time-reversal $\mathcal{T}$ and inversion $\mathcal{P}$ symmetries. Quite differently in 3D AI, $\theta=\pi$ is static, and to the linear order, the magnetic fluctuations has no contributions to the dynamics of axion field~\cite{zhang2019,zhang2020}.

In this Letter, we demonstrate that the out-of-plane surface magnetization dynamics could induce a dynamical magnetoelectric coupling in AI 2D thin films, this further generates a current which is much larger than TME current at magnetic resonance. Interestingly, such a current is finite in the bulk and increases as the film thickness $d$ decreases, which perfectly fits with AI phase of limited $d$ in experiments. The idea can be understood from the response current density by the $\theta$ term,
\begin{equation}\label{response}
\mathbf{j}=\frac{e^2}{2\pi h}\left[\nabla\theta\times\mathbf{E}+\partial_t\theta\mathbf{B}\right],
\end{equation}
where $\partial_t\theta$ could induce a polarization current in gapped systems and can be regarded as a kind of chiral magnetic effect~\cite{fukushima2008,vazifeh2013,zyuzin2012,sekine2016,taguchi2018,armitage2018}. The previous studies on finite-size effect of TME demonstrates $\left(1-\theta(d)/\pi\right)\propto1/d$, where the hybridization between the top and bottom surface states deviates $\theta$ from quantization~\cite{wang2015b,liuzc2020}. We envisaged that the hybridization and thus $\theta$ depend on the surface state exchange gap from the out-of-plane surface magnetization $M_z$ in thin films, which is confirmed by numerical calculations. Therefore $\partial_t\theta$ can be driven by a time-dependent $M_z$ from magnetic resonance. 

\begin{figure}[t]
\begin{center}
\includegraphics[width=3.4in,clip=true]{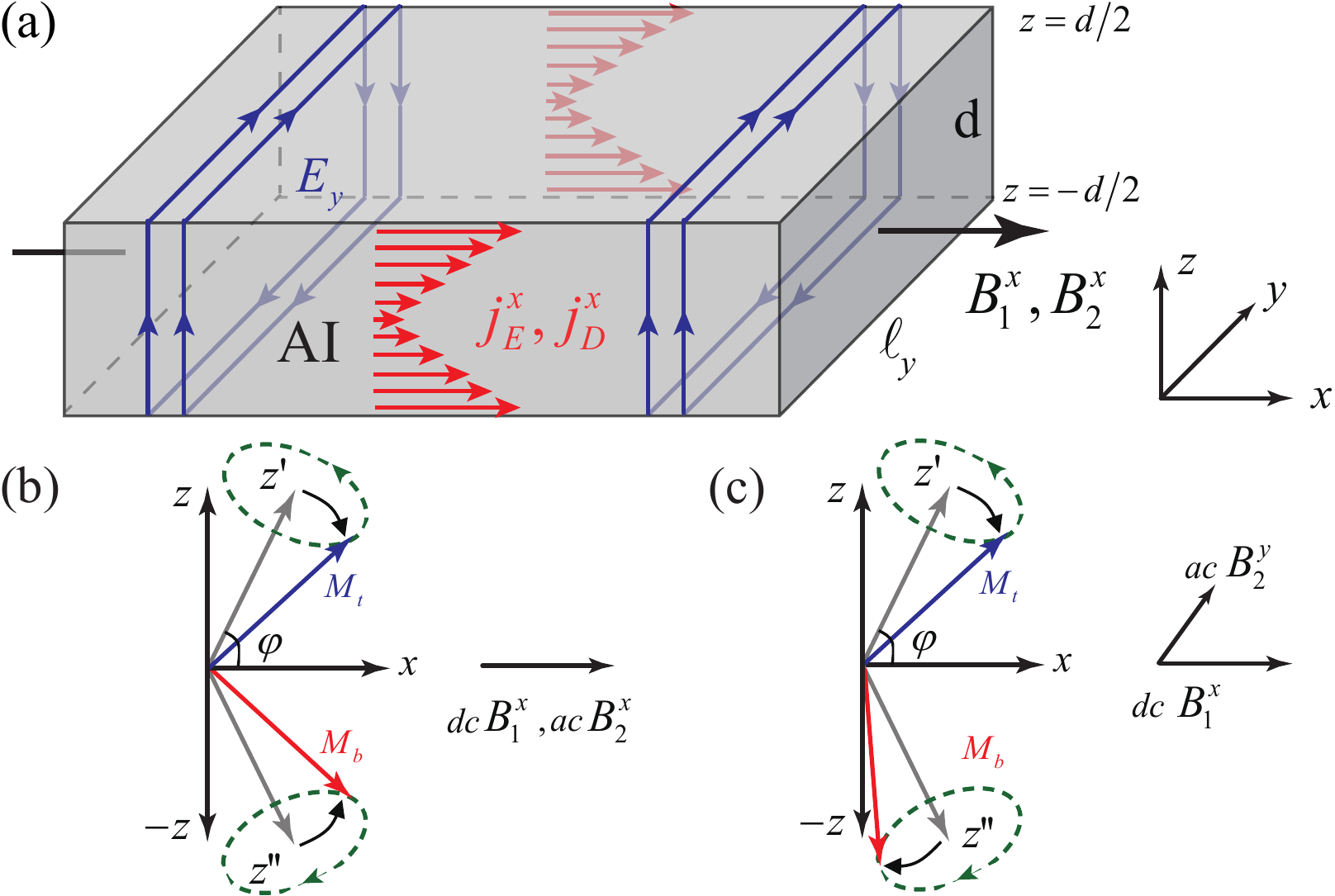}
\end{center}
\caption{(Color online) (a) Illustration of the electric currents $j^x_E$ and $j^x_D$, induced by an ac magnetic field $B^{x}_2\cos(\omega t)$ and a dc field $B^x_1$. $j^x_E$ and $j^x_D$ are induced by $\nabla\theta$ and $\partial_t\theta$, respectively. (b) N\'eel-type and (c) FM-type oscillations from magnetic resonance in different configurations. The antiparallel magnetization along $\pm z$ axis on top and bottom layers now tilt along $z'$ and $z''$.}
\label{fig1}
\end{figure}

\emph{$\theta$ vs $M_z$.} First we examine theoretically the dependence of $\theta$ on the surface $M_z$ in the AI films. For AI films such as FM-TI-FM heterostructure and MnBi$_2$Te$_4$ with a finite thickness along $z$ axis, the linear ME response is diagonal but anisotropic, namely $\alpha_{zz}\neq\alpha_\parallel$~\cite{liuzc2020}. Here $\alpha_{zz}$ and $\alpha_{\parallel}$ are the perpendicular and parallel components of the magnetoelectric susceptibility tensor $\alpha_{ii}$ which relates polarization and magnetic field according to $P_i=-\alpha_{ii}B_i$, and magnetization and electric field correspondingly, $i=x,y,z$. To avoid confusion, we are interested only in the orbital magneoelectric polarizability~\cite{qi2008,essin2009,coh2011} with topological character in $\alpha_{ii}$. The TME response can be directly calculated with the Kubo formula~\cite{wang2015b,liuzc2020} and is defined as the pseudosclar axion part 
\begin{equation}
\frac{\theta}{2\pi}\frac{e^2}{h}=\frac{1}{3}\left(2\alpha_{\parallel}+\alpha_{zz}\right).
\end{equation}

The generic Hamiltonian of AI thin films is written as $\mathcal{H}_{\text{2D}}(\mathbf{k})=\int_{-d/2}^{d/2}dz\mathcal{H}_{\text{3D}}(\mathbf{k},z)$. $\mathbf{k}\equiv(k_x, k_y)$ and we impose periodic boundary conditions in both $x$ and $y$ directions. The physical effect discussed here are generic for any AI thin films and do not rely on a specific model. For concreteness, we adopt the effective Hamiltonian in Ref.~\cite{zhang2019} to describe the low-energy bands of MnBi$_2$Te$_4$ (which is the same for FM-TI-FM heterostructure). The material consists of Van der Waals coupled septuple layers (SL) and develops $A$-type antiferromagnetic (AFM) order with an out-of-plane easy axis, which is FM within each SL but AFM between adjacent SL along $z$ axis. The $\theta=\pi$ in bulk MnBi$_2$Te$_4$ is protected by $\mathcal{P}$ and a combined symmetry $\mathcal{S}\equiv\mathcal{T}\tau_{1/2}$, where $\tau_{1/2}$ is the half translation operator along $z$ axis. In even SL film, $\mathcal{T}, \mathcal{P}$ are broken and thus $\theta\neq\pi$. $\mathcal{H}_{\mathrm{3D}}(\mathbf{k},z)=\varepsilon 1\otimes1+d^1\tau_1\otimes\sigma_2-d^2\tau_1\otimes\sigma_1+d^3\tau_3\otimes1-\Delta(z)1\otimes\sigma_3-iA_1\partial_z\tau_1\otimes\sigma_3$. Here $\tau_j$ and $\sigma_j$ ($j=1,2,3$) are Pauli matrices, $\varepsilon(\mathbf{k},z)=-D_1\partial_z^2+D_2(k_x^2+k_y^2)$, $d^{1,2,3}(\mathbf{k},z)=(A_2k_x, A_2k_y, B_0-B_1\partial_z^2+B_2(k_x^2+k_y^2))$, and $\Delta(z)$ is the $z$-dependent exchange field along $z$ axis. The exchange field in the $xy$ plane will not affect the top and bottom surface gap and thus is neglected here. We assume $\Delta(z)$ takes the values $\pm\Delta_s$ in the top and bottom layers due to antiparallel magnetization, respectively, and zero elsewhere. Explicitly, $\Delta_s =g_MM_z$, where exchange coupling parameter $g_M$ is assumed to be positive and the same on both surfaces for simplicity. All other parameters are taken from Ref.~\cite{zhang2019} for MnBi$_2$Te$_4$ (and similar results in Bi$_2$Te$_3$ family materials).

\begin{figure}[b]
\begin{center}
\includegraphics[width=3.4in,clip=true]{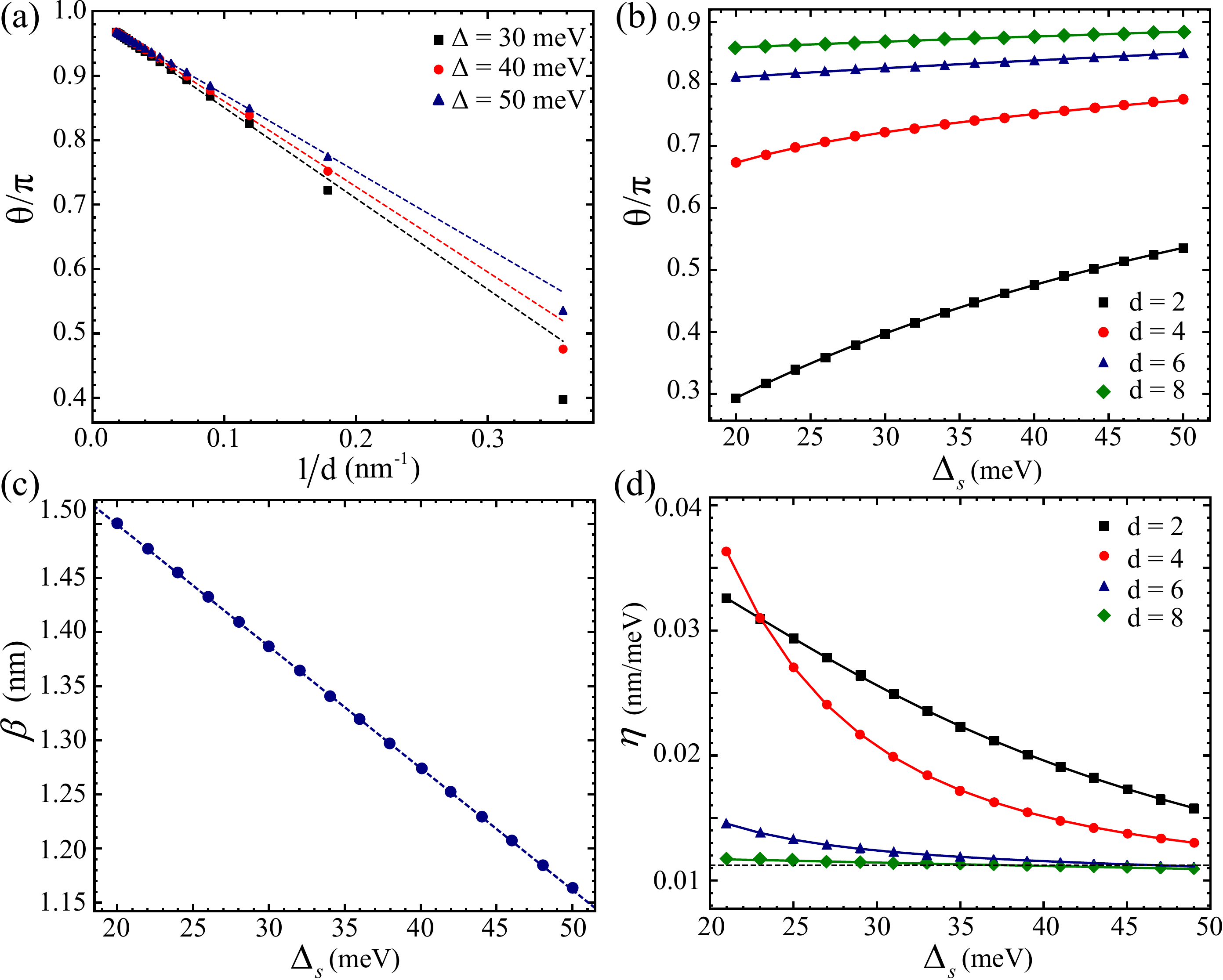}
\end{center}
\caption{(Color online) (a) Finite-size effect of TME. $\theta(d)/\pi$ vs $1/d$ with different typical values of $\Delta_s$. A large deviation from $1/d$ scaling happens when $d$ is small. (b) $\theta/\pi$ vs $\Delta_s$ for different thickness 2, 4, 6, 8 SL. (c) $\beta$ vs $\Delta_s$. (d) $\eta$ vs $\Delta_s$. The value of dashed line is $-\partial\beta/\partial\Delta_s$.}
\label{fig2}
\end{figure}

Fig.~\ref{fig2}(a) shows the numerical calculations of $\theta(d)$ as a function of $1/d$ for different values of $\Delta_s$. The value of $1-\theta(d)/\pi$ scales linearly with $1/d$ as the thickness $d\rightarrow\infty$, while the coefficients depends on $\Delta_s$, namely
\begin{equation}\label{theta}
1-\frac{\theta(d)}{\pi}=\frac{\beta(\Delta_s)}{d}+o\left(\frac{1}{d^2}\right).
\end{equation}
Here $o(1/d^2)$ denotes the higher order terms characterizing the deviation from $1/d$ scaling at small $d$. Fig.~\ref{fig2}(b) shows $\theta(d)$ is a monotonically increasing function of $\Delta_s$ for thin films of 2, 4, 6, and 8 SL. This is consistent with the fact that TME response is from the massive Dirac surface states, and hybridization between the top and bottom surface states partially cancels each others' contributions to TME, which further deviate $\theta$ from quantization. Thus the reduced hybridization from increased $\Delta_s$ will lead $\theta$ closer to quantization. Fig.~\ref{fig2}(c) shows the coefficient $\beta$ decreases linearly as $\Delta_s$ increases. Fig.~\ref{fig2}(d) shows the numerical calculations of $\eta\equiv(d/\pi)(\partial\theta/\partial\Delta_s)$ as a function of $\Delta_s$. The value of $\eta$ deviates from $-\partial\beta/\partial\Delta_s$ for thin films charactering the contribution from $o(1/d^2)$ term in Eq.~(\ref{theta}), and $\eta\rightarrow-\partial\beta/\partial\Delta_s$ quickly converges for thick films such as 8 SL. From Eq.~(\ref{theta}), we get
\begin{equation}\label{dynamical_theta}
\partial_t\theta(d)=\pi\eta g_M\partial_tM_z/d.
\end{equation}
Therefore, $\partial_t\theta$ indeed can be driven by the dynamics of $M_z$ on surfaces, but it vanishes when $d\rightarrow\infty$. Eq.~(\ref{dynamical_theta}) implicitly requires the top and bottom surfaces having opposite $M_z$. Here we point out that the dynamical magnetoelectric coupling in AI films is due to the finite-size effect and vanishes in the bulk, it is essentially different from that in topological AFM materials caused by the bulk magnetic fluctuations which is finite as $d\rightarrow\infty$~\cite{li2010}. 

\emph{$j_E$ vs $j_D$.} Now we study the response current from the spatial and temporal gradient of $\theta$ in Eq.~(\ref{response}). They are the two sides of same coin demonstrating TME. Considering the process of applying a uniform external dc magnetic field $B^x_1\hat{\mathbf{x}}$ and ac field $B^x_2\cos(\omega t)\hat{\mathbf{x}}$ of frequency $\omega/2\pi$ in Fig.~\ref{fig1}(a). The oscillating $B^x_2$ can induce a non-uniform electric field along $y$ due to the Faraday’s law: $\mathbf{E}(t,z)=-\omega B^x_2\sin(\omega t)z\hat{\mathbf{y}}$ with $z=0$ set at the middle of the AI layer. From the first term in Eq.~(\ref{response}), this further induces a Hall current density $j^x_E=(\partial_z\theta/2\pi)(e^2/h)\hat{\mathbf{z}}\times\mathbf{E}$. Thus the integration over $z$ gives the TME current density in 2D
\begin{equation}
\mathcal{J}_E=\mathcal{J}^x_E\hat{\mathbf{x}}=(\theta/2\pi)(e^2/h)\omega dB^x_2\sin(\omega t)\hat{\mathbf{x}},
\end{equation}
whose amplitude is proportional to $\theta$ and limited by $d$. Here $d$ is maximally $10$~nm in experiments to ensure the full insulating state~\cite{mogi2017,mogi2017a,xiao2018,liu2020,deng2020}. For an estimation, with typical parameters $B^x_2=5$~G, $\omega/2\pi=7$~GHz, $d=5.6$~nm, $\theta/\pi\approx0.7$ (finite-size effect taken into account as in Fig.~\ref{fig2}), and $\ell_y=500$~$\mu$m (the length of film along $y$), the amplitude of TME current is $I_E^x=|\text{max}(\mathcal{J}^x_E)|\ell_y=0.83$~nA, which is quite small to be measured. 

\begin{figure}[t]
\begin{center}
\includegraphics[width=3.4in,clip=true]{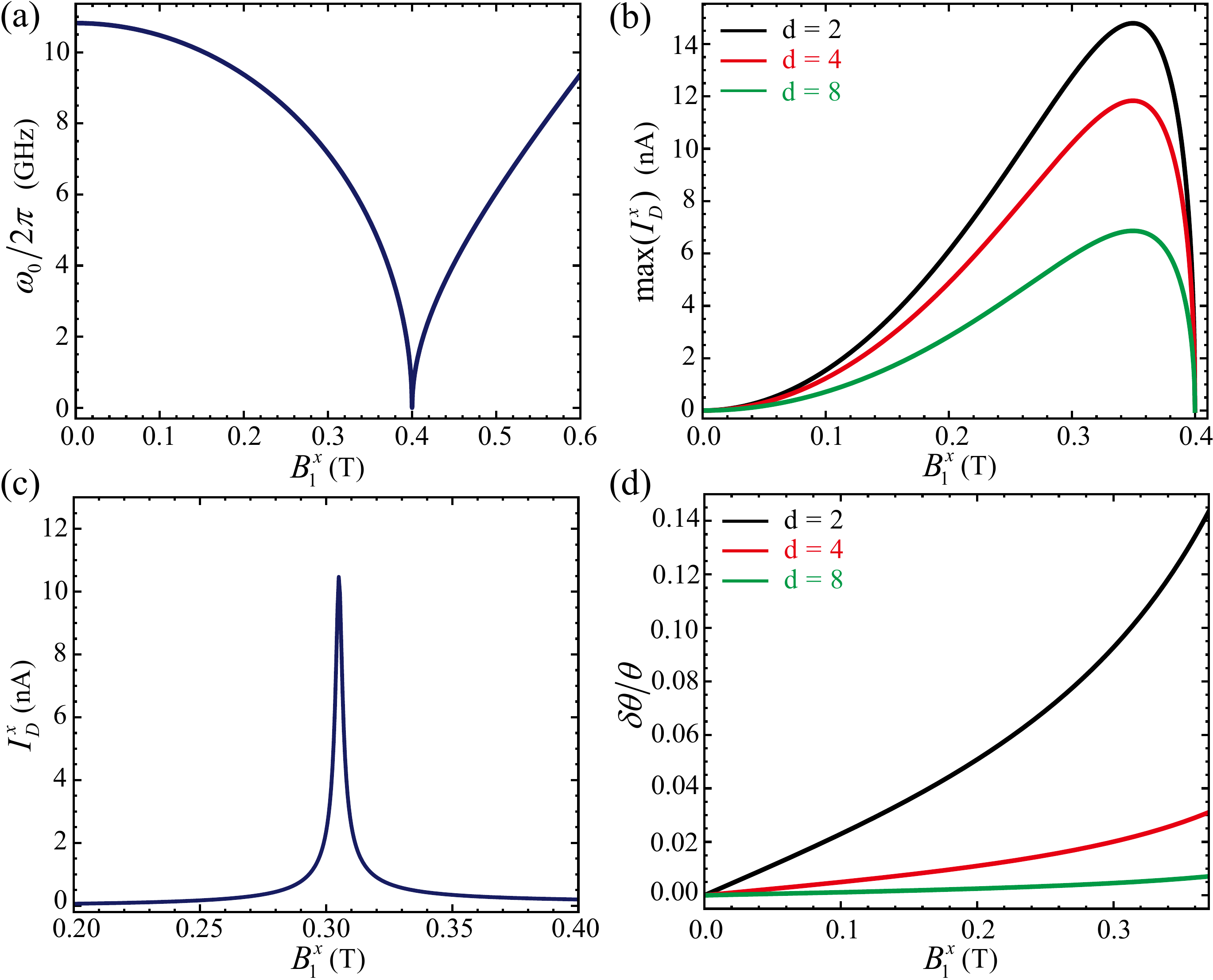}
\end{center}
\caption{(Color online) FMR. (a) The FMR frequency $\omega_0/2\pi$ vs $B^x_1$. (b) The resonant amplitude of $I_D^x=|\text{max}(\mathcal{J}^x_D)|\ell_y$ at FMR vs $B^x_1$ for 2, 4, and 8 SL. (c) The response of $I_D^x$ amplitude as a function of scanning $B^x_1$ for 4 SL, with the microwave field frequency fixed at $\omega/2\pi=7$~GHz. (d) $\delta\theta/\theta$ vs $B^x_1$.
Here $K=2\times10^4$~J/m$^3$, $M_s=10^5$~A/m, $\alpha=5\times10^{-3}$.}
\label{fig3}
\end{figure}

Meanwhile, the surface magnetic moments are tilted away from $\pm z$ axis by $B^x_1$, and $B^x_2$ induces an oscillating $M_z$ with a same frequency. One can decompose $M_z$ into the static and dynamical parts as $M_z=M_0^z+\delta M_z(t)$. In this configuration as shown in Fig.~\ref{fig1}(a), the top and bottom surfaces have opposite $\delta M_z(t)$, which can be dubbed as N\'eel-type oscillation. Thus the 2D current density $\mathcal{J}_D$ induced by $\partial_t\theta$ is
\begin{equation}
\mathcal{J}_D = \mathcal{J}_D^x\hat{\mathbf{x}} = \frac{e^2}{2h}\eta g_M\partial_t\delta M_z\left(B^x_1+B^x_2\cos(\omega t)\right)\hat{\mathbf{x}},
\end{equation}
whose amplitude is proportional to $\eta$ and thus increases as $d$ decreases, in sharp contrast to $\mathcal{J}_E$ which decreases as $d$ decreases. Strikingly, even though $\partial_t\theta$ vanishes when $d\rightarrow\infty$, the induced 2D current density $\mathcal{J}_D$ is finite in the bulk and is \emph{independent} of $d$ when $d$ is large enough. There are $\omega$ and $2\omega$ components in $\mathcal{J}_D^x$, which are proportional to $B_1^x$ and $B_2^x$, respectively. In particular, for finite films, the $2\omega$ component $\mathcal{J}_D^x(2\omega)/\mathcal{J}^x_E=\delta\theta/\theta<0.1$, with $\delta\theta\equiv(\partial\theta/\partial\Delta_s)g_M\delta M_z$ the oscillating part in $\theta$ due to $\delta M_z$, which is maximized at the resonant frequency of the FM layer. $\delta\theta/\theta$ is on the order of $0.01\sim0.1$ as calculated in Fig.~\ref{fig3}(d). Therefore, $\mathcal{J}^x_D(2\omega)$ can be neglected. In the following we focus on only $\mathcal{J}_D^x(\omega)$.

\emph{FMR induced $\delta M_z$.} First we consider the dynamics of $M_z$ induced by FM resonance (FMR) in FM-TI-FM heterostructure. The tilted magnetization is along $z'$ ($z''$), where the angle between $z'$ ($z''$) and $x$ axis is $\varphi$ as shown in Fig.~\ref{fig1}(b). $\cos\varphi=B^x_1M_s/2K$, $M_s=|\mathbf{M}|$ is the saturation magnetization, $K$ is the effective uniaxial anisotropy. Here we consider $\varphi\neq0$ to ensure it is always in AI phase. The two FM layers are decoupled, and the magnetization dynamics governed by the Landau-Lifshitz-Gilbert (LLG) equations~\cite{kittel1948,mattis1988} for two FM layers under the same $B^x_2(t)$ have the same form. For simplicity, we assume the damping constant and $M_s$ are the same in two FM layers. The equation can be solved by linearization~\cite{kittel1948,supplementary}, and the steady solution of $\delta M_z$ at FMR is given by
\begin{equation}\label{deltaMz}
\delta M_z = \frac{\gamma\omega_1B^x_2M_s\sin2\varphi}{2\alpha\omega_0\left(2\omega_1-\gamma B^x_1\cos\varphi\right)}\sin(\omega_0 t),
\end{equation}
where $\gamma=\gamma_0/(1+\alpha^2)$, $\gamma_0=2e/(2m_e)$ is the gyromagnetic ratio of an electron, $\alpha$ is dimensionless Gilbert damping constant, $\omega_0=\sqrt{\omega_1(\omega_1-\gamma B^x_1\cos\varphi)}$ is resonance frequency, $\omega_1=\gamma(B^x_1\cos\varphi+2K\sin^2\varphi/M_s)$. Obviously, $\delta M_z\neq0$ when $\varphi\neq\pi/2$. The adiabatic approximation always holds, for the energy scale of the typical FMR frequency range $\omega_0/2\pi=1\sim10$~GHz is much smaller than the surface magnetic gap. Then the 2D current density $\mathcal{J}_D^x$ at FMR is
\begin{equation}
\mathcal{J}_{D}^x = \frac{e^2}{2h}\frac{\gamma B^x_1B^x_2\eta g_MM_s\omega_1\sin2\varphi}{2\alpha\left(2\omega_1-\gamma B_x\cos\varphi\right)}\cos(\omega_0 t).
\end{equation}
With a fixed microwave frequency $\omega$, one can scan the field strength of $B^x_1$ to achieve FMR. The resonant frequency of the FM layer $\omega_0$ versus $B^x_1$ is calculated in Fig.~\ref{fig3}(a), where $\omega_0=0$ represents the magnetization is just tuned to be in-plane, namely $\varphi=0$. With similar typical parameters $B^x_2=5$~G, $d=5.6$~nm, $\ell_y=500$~$\mu$m, and $\alpha=5\times10^{-3}$ in FM~\cite{wu2020}, then the estimated amplitude of $I_D^x$ versus $B^x_1$ is shown in Fig.~\ref{fig3}(b), where the maximum value is about $12$~nA, in the range accessible by transport experiments. 

We compare the ratio between the amplitudes of $\mathcal{J}_D^x$ and $\mathcal{J}_E^x$ as $\mathcal{R}\equiv\left|\text{max}(\mathcal{J}_D^x)/\text{max}(\mathcal{J}_E^x)\right|=(\delta\theta/\theta)(B^x_1/B^x_2)$. With $\delta\theta/\theta\approx0.01\sim0.1$, and $B^x_1=0.1\sim0.4$~T, the ratio is approximately $\mathcal{R}\approx10^1\sim10^2$. Thus the current induced by magnetic dynamics at FMR is the dominant contribution. Importantly, it is larger in thin film than that in thick one, which fits well with the experimental condition of limited $d$. Moreover, TME vanishes for the thin films of trivial insulating states (bulk $\theta=0$) without topological surface states. Therefore, $I_D^x$ can be used to distinguish the AI from a trivial insulator experimentally.

\begin{figure}[t]
\begin{center}
\includegraphics[width=3.4in,clip=true]{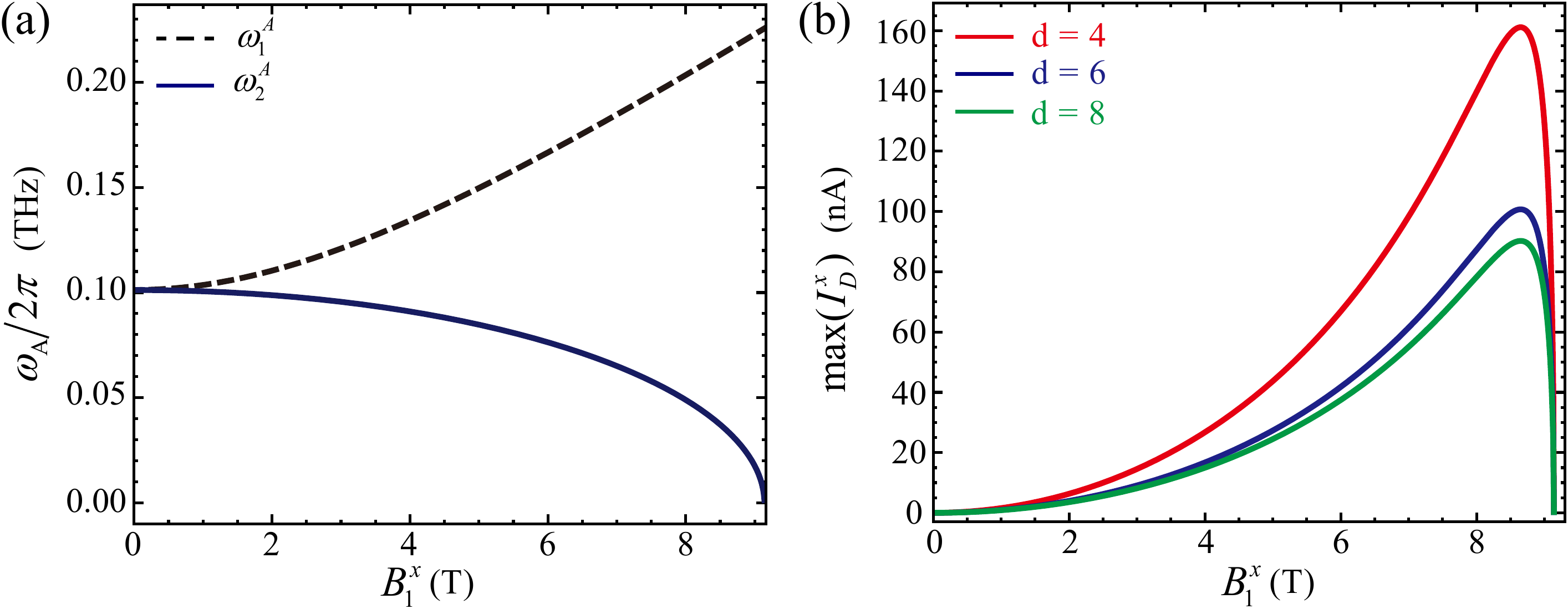}
\end{center}
\caption{(Color online) AFMR. (a) Two branches of AFMR frequency $\omega_A/2\pi$ vs $B^x_1$. (b) The amplitude of $I_D^x$ at AFMR in the $\omega^A_2$ branch vs $B^x_1$ for 4, 6, and 8 SL, and that in the $\omega^A_1$ branch almost vanishes (not shown).}
\label{fig4}
\end{figure}

\emph{AFMR induced $\delta M_z$.} Then we study $\partial_tM_z$ induced by AFM resonance in AI such as MnBi$_2$Te$_4$. 
This is the the simplest bipartite collinear AFM, where the magnetic dynamics of the surface $M_z$ is governed by the LLG equations by including the exchanging coupling term between neighboring SL due to the intrinsic magnetism. The tilted magnetization is along $z'$ ($z''$) with the angle $\varphi$ between $z'$ ($z''$) and $x$, but now $\cos\varphi=B^x_1/(4B_E+2B_A)$, $B_E\equiv J_AM_s$ and $B_A\equiv K_1/M_s$ are the exchange field and anisotropy field, respectively. The equations can be solved by linearization and numerically~\cite{supplementary}. For an estimation, take exchange coupling $J_A=0.55$~meV, effective anisotropy field $K_1=0.22$~meV~\cite{otrokov2019,lib2020}, $\alpha=5\times10^{-3}$, $M_s=2\times10^5$~A/m, $B^x_2=5$~G, the AFMR frequency $\omega_A$ vs $B^x_1$ is shown in Fig.~\ref{fig4}(a). The two branches $\omega^A_1$ and $\omega^A_2$ represents the resonance from AFM and FM components, respectively. The estimated $I^x_D$ amplitude is calculated in Fig.~\ref{fig4}(b), where the maximum value is about $100$~nA in $\omega^A_2$ branch, and is negligible in $\omega^A_1$ branch. Then the ratio between $\mathcal{J}_D^x$ and $\mathcal{J}_E^x$ is about $\mathcal{R}\approx1\sim20$. Therefore, the dynamical current at AFMR is about one order of magnitude larger than TME current.

\emph{Different configuration.} Then we discuss a different configuration where the ac magnetic field $B_2^y$ is applied along $y$ axis, but keeping the dc field $B_1^x$ along $x$ axis. The top and bottom surfaces now have the same oscillating $\delta M_z(t)$ induced by $B^y_2$ for both FMR and AFMR~\cite{supplementary}, and can be dubbed as FM-type oscillation. To quantify how the same $\delta M_z(t)$ affect $\delta\theta$, we calculate $\theta$ versus $\delta\Delta_s$, where $\Delta(z)$ takes the value $\Delta_s+\delta\Delta_s$ and $-\Delta_s+\delta\Delta_s$ on the top and bottom layers, respectively. We find $\theta$ almost unchanged by varying $\delta\Delta_s$, specifically, $\delta\theta/\theta$ is $10^{-3}$ smaller compared to that from N\'eel-type oscillation~\cite{supplementary}. Therefore, $\mathcal{J}_D^x$ almost vanishes compared to $\mathcal{J}_E^y$ in the new configuration, which provides another testable signature for our theory.

\emph{Discussion.} We have demonstrated an intimate relations between surface magnetization dynamics and dynamical magnetoelectric coupling in AI thin films, which could further generate a measurable polarization current but is absent in trivial insulators. Our theory is fundamentally different from the pseudo-electric field induced current discussed in Ref.~\cite{yu2019}. In Ref.~\cite{yu2019}, the current is from the first term in Eq.~(\ref{response}), where pseudo-electric field is induced by in-plane magnetization dynamics and is maximized when magnetization is oscillating around $z$ axis; while in our case, the current is from the second term in Eq.~(\ref{response}), where $\partial_t\theta$ is driven by the out-of-plane surface magnetization dynamics and is maximized when magnetization is tiled away from $z$ axis. Also, the current in Ref.~\cite{yu2019} is proportional to $\theta$, which decreases as $d$ decreases similar to TME current; while thickness dependence of the current in our case is just the opposite, namely, the current increases as $d$ decreases. Moreover, the dynamical magnetoelectric coupling  in AI thin films is from the finite-size effect, which cannot exist in trivial insulators. It is also essentially different from the dynamical axion field induced by magnetic fluctuations~\cite{li2010}, which can exist in 3D $\mathcal{T,P}$-broken insulators, regardless of topological or trivial. 

It is worth mentioning that if the top and bottom layers has opposite exchange coupling parameters, only parallel magnetization realizes AI. Then the dynamical current is largest from FM-type oscillation, and almost vanishes due to N\'eel-type oscillation. Recently, FMR has been realized in FM-TI heterostructure~\cite{wu2020}, together with the experimental observation of zero Hall plateau in FM-TI-FM heterostructure~\cite{mogi2017,mogi2017a,xiao2018} and MnBi$_2$Te$_4$ even SL~\cite{liu2020,deng2020}, making the realization of the dynamical magnetoelectric current predicted here in AI films feasible. 

\begin{acknowledgments}
We acknowledge B. Lian, Y. Wu, and Y. Wang for valuable discussions. This work is supported by the National Key Research Program of China under Grant Nos.~2016YFA0300703 and 2019YFA0308404, the Natural Science Foundation of China through Grant Nos.~11774065 and 11722430, Shanghai Municipal Science and Technology Major Project under Grant No.~2019SHZDZX01, and the Natural Science Foundation of Shanghai under Grant No.~19ZR1471400.
\end{acknowledgments}

\end{document}